\newcommand{\diracslash}[1]{#1\llap{/\kern2pt}}

\newcommand{\be}{\begin{equation}}
\newcommand{\ee}{\end{equation}}
\newcommand{\bea}{\begin{eqnarray}}
\newcommand{\eea}{\end{eqnarray}}
\newcommand{\ba}[1]{\begin{array}{#1}}
\newcommand{\ea}{\end{array}}

\newcommand{\bt}{\begin{tabular}}
\newcommand{\et}{\end{tabular}}

\newcommand{\beas}{\begin{eqnarray*}}
\newcommand{\eeas}{\end{eqnarray*}}

\documentclass[preprint,prd,aps,floats,nofootinbib,floatfix]{revtex4}
\usepackage{graphicx}
\begin{document}

\title{Upsilon states in magnetized nuclear matter}
\author{Amal Jahan C.S.}
\email{amaljahan@gmail.com}
\affiliation{Department of Physics, Indian Institute of Technology, Delhi,
Hauz Khas, New Delhi -- 110 016, India}

\author{Shivam Kesarwani}
\email{kesarishvam@gmail.com}
\affiliation{Department of Physics, Indian Institute of Technology, Delhi,
Hauz Khas, New Delhi -- 110 016, India}

\author{Sushruth Reddy P.}
\email{sushruth.p7@gmail.com}
\affiliation{Department of Physics, Indian Institute of Technology, Delhi,
Hauz Khas, New Delhi -- 110 016, India}

\author{Nikhil Dhale}
\email{dhalenikhil07@gmail.com}
\affiliation{Department of Physics, Indian Institute of Technology, Delhi,
Hauz Khas, New Delhi -- 110 016, India}

\author{Amruta Mishra}
\email{amruta@physics.iitd.ac.in}
\affiliation{Department of Physics, Indian Institute of Technology, Delhi,
Hauz Khas, New Delhi -- 110 016, India}

\begin{abstract}
The mass modifications of the bottomonium states
($\Upsilon (NS)$,$N=1,2,3,4$ and $\Upsilon (1D)$)
in magnetized nuclear matter are studied using 
chiral effective model. The in-medium masses are 
calculated from the medium modification of the scalar 
dilaton field in the chiral effective model, 
which simulates the gluon condensates of QCD. 
The strengths of the wave functions (assumed 
to be harmonic oscillator wave functions) denoted 
by the parameter, $\beta$, of the $\Upsilon (NS)$, 
$N=1,2,3,4$, are fitted from their observed decay 
widths to $e^+e^-$. The decay width for the channel,
$\Upsilon (1D) \rightarrow e^+ e^-$ 
yet been observed experimentally, has been predicted 
in the present work, by using the value of the parameter, 
$\beta$ for $\Upsilon(1D)$, interpolated from the $\beta$ 
versus mass relation, for the upsilon states.
The effects of the isospin asymmetry of the nuclear medium
on the masses of the upsilon states are investigated 
and are observed to be large for high densities. 
This should have observable consequences at the asymmetric 
heavy ion collisions at the Compressed baryonic matter (CBM) 
experiments at FAIR, GSI as well as at SPS, CERN. 
The study of the bottomonium states at CBM will however require
access to higher energies than the energy regime
planned at present. The effects of magnetic field on the 
masses of bottomonium states in nuclear matter
are studied in the present work. These masses
are investigated including the anomalous magnetic moments (AMM)
of the nucleons, and compared to the results
when the AMMs of nucleons are not taken into account.
The effects of magnetic field as well as isospin asymmetry
on the upsilon masses are obserevd to be large at high densities.
\end{abstract}
\maketitle

\def\bfm#1{\mbox{\boldmath $#1$}}

\section{Introduction}

The study of the in-medium properties of the hadrons
has been a topic of intense research in the recent past,
due to its relevance in ultra-relativistic 
heavy ion collision experiments.
In these experiments, matter at high density and/or temperature
is created and the experimental observables are affected by
modifications of the hadrons in the strongly interacting medium.
In ultra-relativistic heavy ion collision experiments,
e.g., at RHIC at BNL and at LHC at CERN,
large magnetic fields are also believed to be created \cite{HIC_mag}. 
This has initiated a lot of work to study the effects 
of strong magnetic fields on the properties of hadrons
in the medium. The study of heavy flavoured hadrons 
\cite{Hosaka_Prog_Part_Nucl_Phys} has also attracted a lot 
of attention in the recent years, as these are being
(planned to be) investigated extensively in the heavy 
ion collision experiments. Within the QCD sum rule approach,
the masses of the heavy quarkonium states (charmonium and
bottomonium states) are modified in the hadronic medium, 
due to the medium modifications of the gluon condensates
in QCD \cite{kimlee,klingl,leeko,amarvjpsi_qsr}. On the other hand,
the medium modifications of the masses of the light vector mesons 
\cite{hatsuda,am_vecmeson_qsr}, 
as well as of the open charm (bottom) mesons 
\cite{open_heavy_flavour_qsr,Wang_heavy_mesons,arvind_heavy_mesons_QSR}
arise due to the modifications of the light quark condensates.
In the literature, the open heavy flavour mesons have also been studied 
using the coupled channel approach
\cite{ltolos,ljhs,mizutani,HL,tolos_heavy_mesons}, 
the quark meson coupling model \cite{open_heavy_flavour_qmc},
due to pion exchange with nucleons  \cite{Yasui_Sudoh_pion},
the heavy meson effective theory
\cite{Yasui_Sudoh_heavy_meson_Eff_th}, studying the
heavy flavour meson as an impurity in nuclear matter 
\cite{Yasui_Sudoh_heavy_particle_impurity}, as well as,
using a chiral effective model 
\cite{amarindamprc,amarvdmesonTprc,amarvepja,amdmeson,DP_AM_Ds,
DP_AM_bbar,DP_AM_Bs}.
There have also been predictions for the heavy meson-nucleus
bound states \cite{krein_18,krein_jpsi}. 
The formation of such bound states could be possible,
due to the attractive interaction of the heavy mesons 
in the nuclear medium 
\cite{leeko,amarvdmesonTprc,amarvepja,krein_jpsi}.
The huge magnetic fields created in ultra-relativistic
heavy ion collision experiments has also initiated
the study of the heavy flavour mesons
in presence of strong magnetic fields
\cite{Gubler_D_mag_QSR,B_mag_QSR,charmonium_mag_QSR,
dmeson_mag,bmeson_mag,charmonium_mag}.

We study the mass modifications of the upsilon states 
($\Upsilon (NS)$, $N=1,2,3,4$ and $\Upsilon (1D))$ 
arising due to medium change of a scalar dilaton field,
within a chiral effective model \cite{Schechter,paper3,kristof1},
which is incorporated in the model to simulate 
the gluon condensates of QCD. 
The chiral effective model has been used 
to study finite nuclei \cite{paper3},
hot hyperonic matter \cite{kristof1}, 
in-medium properties of the light vector 
($\omega$, $\rho$ and $\phi$) mesons \cite{hartree}, 
and the kaons and antikaons
\cite{kaon_antikaon,isoamss,isoamss1,isoamss2}
as well as to investigate the bulk matter
in the interior of (proto) neutron stars 
\cite{pneutronstar}.
The chiral SU(3) model has also been generalized to 
SU(4) as well as to SU(5) to derive the interactions 
of the charm as well as bottom mesons with the light 
hadronic sector. Using these interactions, 
the open (strange) charm mesons
\cite{amarindamprc,amarvdmesonTprc,amarvepja,amdmeson,DP_AM_Ds},
open (strange) bottom mesons \cite{DP_AM_bbar,DP_AM_Bs}, 
the charmonium states \cite{amarvdmesonTprc,amarvepja}, 
the upsilon states \cite{AM_DP_upsilon} have been studied. 
Within the chiral effective model, the mass modifications
of the open charm (bottom) mesons arise due to their interactions
with the baryons and scalar mesons, whereas the mass
modifications of the charmonium and bottomonium states
arise due to interaction with the scalar dilaton field,
which simulates the gluon condensates of QCD.
Using the medium modifications of the 
charmonium states as well as the $D$ and $\bar D$ mesons
as calculated within the chiral effective model,
the partial decay widths
of the charmonium states to $D\bar D$ in the hadronic medium
\cite{amarvepja} have been studied 
using a light quark creation model \cite{friman},
namely the $^3P_0$ model \cite{3p0}. The in-medium
decay widths of the charmonium (bottomonium) to $D\bar D$
($B\bar B$) have also been investigated using a field theoretic model
for composite hadrons \cite{amspmwg,amspm_upsilon}.
Recently, the effects of magnetic fields on the masses 
of the open charm $D$ and $\bar D$ mesons 
\cite{dmeson_mag}, the open bottom mesons \cite{bmeson_mag},
as well as, the charmonium states \cite{charmonium_mag} 
in asymmetric nuclear matter,
have also been studied using the chiral effective model. 

The outline of the paper is as follows : In section II, we describe
briefly the chiral effective model used to investigate the
medium modifications of the upsilon states in
magnetized nuclear matter. These modifications arise
due to medium change of the dilaton field, which mimics the
gluon condensates of QCD.
The effects of the anomalous magnetic moments of the
nucleons on the bottomonium masses are also investigated
for different values of the magnetic field.
In section III, we discuss the results obtained 
for the in-medium masses of the bottomonium states
in strong magnetic fields. In section IV, we summarize 
the findings of the present study.

\section{In-medium masses of the bottomonium states}

The medium modifications of the bottomonium masses
in strongly magnetized asymmetric nuclear matter
are studied using a chiral effective model \cite{paper3}.
The model is based on the nonlinear realization of chiral 
symmetry \cite{weinberg, coleman, bardeen} and broken scale invariance 
\cite{paper3,kristof1,hartree}. 
The effective hadronic chiral Lagrangian density contains the following terms
\be
{\cal L} = {\cal L}_{kin} + {\cal L}_{BM}
          + {\cal L}_{vec} + {\cal L}_0 +
{\cal L}_{scalebreak}+ {\cal L}_{SB}+{\cal L}_{mag}.
\label{genlag} \ee 
In the above Lagrangian density, the first term ${\cal L}_{kin}$ 
corresponds to the kinetic energy terms of the baryons and the mesons.
${\cal L}_{BM}$ is the baryon-meson interaction term,
${\cal L}_{vec}$  corresponds to the interactions of the vector 
mesons, ${\cal L}_{0}$ contains the meson-meson interaction terms, 
${\cal L}_{scalebreak}$ is a scale invariance breaking logarithmic 
potential given in terms of a scalar dilaton field \cite{heide1}, 
$ {\cal L}_{SB} $ is the explicit chiral symmetry
breaking term, and ${\cal L}_{mag}$ is the contribution 
from the magnetic field, given as 
\cite{broderick1,broderick2,Wei,mao,dmeson_mag,bmeson_mag,charmonium_mag}
\be 
{\cal L}_{mag}=-{\bar {\psi_i}}q_i \gamma_\mu A^\mu \psi_i
-\frac {1}{4} \kappa_i \mu_N {\bar {\psi_i}} \sigma ^{\mu \nu}F_{\mu \nu}
\psi_i
-\frac{1}{4} F^{\mu \nu} F_{\mu \nu},
\label{lmag}
\ee
where, $\psi_i$ is the field operator for the $i$-th baryon 
($i=p,n$, for nuclear matter, as considered in the present work),
and the parameter $\kappa_i$ in the second term in equation (\ref{lmag}) 
is related to the anomalous magnetic moment of the $i$-th baryon
\cite{broderick1,broderick2,Wei,mao,amm,VD_SS,aguirre_fermion}.
The values of $\kappa_p$ and  $\kappa_n$ are given as 
$3.5856$ and $-3.8263$ respectively, which are the values
of the gyromagnetic ratio corresponding to the 
anomalous magnetic moments of the proton and 
neutron respectively.

In the present study of the in-medium upsilon masses
in magnetized nuclear matter using the chiral SU(3) model, 
we use the mean field approximation,
where all the meson fields are treated as classical fields. 
The coupled equations of motion for the non-strange scalar field $\sigma$, 
strange scalar field $ \zeta$, scalar-isovector field $ \delta$ and dilaton 
field $\chi$, are derived from the Lagrangian density
and are given as
\begin{eqnarray}
&& k_{0}\chi^{2}\sigma-4k_{1}\left( \sigma^{2}+\zeta^{2}
+\delta^{2}\right)\sigma-2k_{2}\left( \sigma^{3}+3\sigma\delta^{2}\right)
-2k_{3}\chi\sigma\zeta \nonumber\\
&-&\frac{d}{3} \chi^{4} \bigg (\frac{2\sigma}{\sigma^{2}-\delta^{2}}\bigg )
+\left( \frac{\chi}{\chi_{0}}\right) ^{2}m_{\pi}^{2}f_{\pi}
-\sum g_{\sigma i}\rho_{i}^{s} = 0 
\label{sigma}
\end{eqnarray}
\begin{eqnarray}
&& k_{0}\chi^{2}\zeta-4k_{1}\left( \sigma^{2}+\zeta^{2}+\delta^{2}\right)
\zeta-4k_{2}\zeta^{3}-k_{3}\chi\left( \sigma^{2}-\delta^{2}\right)\nonumber\\
&-&\frac{d}{3}\frac{\chi^{4}}{\zeta}+\left(\frac{\chi}{\chi_{0}} \right) 
^{2}\left[ \sqrt{2}m_{k}^{2}f_{k}-\frac{1}{\sqrt{2}} m_{\pi}^{2}f_{\pi}\right]
 -\sum g_{\zeta i}\rho_{i}^{s} = 0 
\label{zeta}
\end{eqnarray}
\begin{eqnarray}
& & k_{0}\chi^{2}\delta-4k_{1}\left( \sigma^{2}+\zeta^{2}+\delta^{2}\right)
\delta-2k_{2}\left( \delta^{3}+3\sigma^{2}\delta\right) +2k_{3}\chi\delta 
\zeta \nonumber\\
& + &  \frac{2}{3} d \chi^{4} \left( \frac{\delta}{\sigma^{2}-\delta^{2}}\right)
-\sum g_{\delta i}\rho_{i}^{s} = 0
\label{delta}
\end{eqnarray}
\begin{eqnarray}
& & k_{0}\chi \left( \sigma^{2}+\zeta^{2}+\delta^{2}\right)-k_{3}
\left( \sigma^{2}-\delta^{2}\right)\zeta + \chi^{3}\left[1
+{\rm {ln}}\left( \frac{\chi^{4}}{\chi_{0}^{4}}\right)  \right]
+4k_{4}\chi^{3}
\nonumber\\
& - & \frac{4}{3} d \chi^{3} {\rm {ln}} \Bigg ( \bigg (\frac{\left( \sigma^{2}
-\delta^{2}\right) \zeta}{\sigma_{0}^{2}\zeta_{0}} \bigg ) 
\bigg (\frac{\chi}{\chi_0}\bigg)^3 \Bigg ) 
+\frac{2\chi}{\chi_{0}^{2}}\left[ m_{\pi}^{2}
f_{\pi}\sigma +\left(\sqrt{2}m_{k}^{2}f_{k}-\frac{1}{\sqrt{2}}
m_{\pi}^{2}f_{\pi} \right) \zeta\right]  = 0 
\label{chi}
\end{eqnarray}
In the above, ${\rho_i}^s (i=p,n)$ are the scalar densities for the nucleons. 
%
%
In the presence of magnetic field, the proton has contributions 
from the Landau energy levels. The number density and the 
scalar density of the proton are given as \cite{Wei,mao}

\begin{equation}
\rho_p=\frac{eB}{4\pi^2} \Bigg [ 
\sum_{\nu=0}^{\nu_{max}^{(s=1)}} k_{f,\nu,1}^{(p)} 
+\sum_{\nu=1}^{\nu_{(max)}^{(s=-1)}} k_{f,\nu,-1}^{(p)} 
\Bigg]
\end{equation}
and 
\begin{eqnarray}
\rho_s^p & = & \frac{eB{m_p}^*}{2\pi^2} \Bigg [ 
\sum_{\nu=0}^{\nu_{max}^{(s=1)}}
\frac {\sqrt {{m_p^*}^2+2eB\nu}+\Delta_p}{\sqrt {{m_p^*}^2+2eB\nu}}
\ln |\frac{ k_{f,\nu,1}^{(p)} + E_f^{(p)}}{\sqrt {{m_p^*}^2
+2eB\nu}+\Delta_p}|\nonumber \\
 &+&\sum_{\nu=1}^{\nu_{max}^{(s=-1)}}
\frac {\sqrt {{m_p^*}^2+2eB\nu}-\Delta_p}{\sqrt {{m_p^*}^2+2eB\nu}}
\ln |\frac{ k_{f,\nu,-1}^{(p)} + E_f^{(p)}}{\sqrt {{m_p^*}^2
+2eB\nu}-\Delta_p}|\Bigg ]
\end{eqnarray}
where, $k_{f,\nu,\pm 1}^{(p)}$ are the Fermi momenta of protons
for the Landau level, $\nu$ for the spin index, $s=\pm 1$,
i.e. for spin up and spin down projections for the proton.
These Fermi momenta are related to the Fermi energy of the
proton as
\begin{equation}
k_{f,\nu,s}^{(p)}=\sqrt { {E_f^{(p)}}^2
-\Big (
{\sqrt {{m_p^*}^2+2eB\nu}+s\Delta_p}\Big )^2}.
\end{equation}
The number density and the scalar density of neutrons are given as
\begin{equation}
\rho_{n}= \frac{1}{4\pi^2} \sum _{s=\pm 1}
\Bigg \{ \frac{2}{3} {k_{f,s}^{(n)}}^3
+s\Delta_n \Bigg[ (m_n^*+s\Delta_n) k_{f,s}^{(n)}
+{E_f^{(n)}}^2 \Bigg( arcsin \Big (
\frac{m_n^*+s\Delta_n}{E_f^{(n)}}\Big)-\frac{\pi}{2}\Bigg)\Bigg]
\Bigg \}
\end{equation}
and
\begin{equation}
\rho_s^n =\frac{m_n^*}{4\pi^2} \sum _{s=\pm 1} 
\Bigg [ k_{f,s}^{(n)} E_f^{(n)} - 
(m_n^*+s\Delta_n)^2 \ln | \frac {k_{f,s}^{(n)}+ 
E_f^{(n)}}{m_n^*+s\Delta_n} | \Bigg]
\end{equation}

In the above, the Fermi momentum, $k_{f,s}^{(n)}$ 
for the neutron with spin projection, s 
($s=\pm 1$ for the up (down) spin projection), 
is related to the Fermi energy for the 
neutron, $E_f^{(n)}$ as
\begin{equation}
k_{f,s}^{(n)}= \sqrt { {E_f^{(n)}}^2 -
(m_n^*+s\Delta_n)^2},
\end{equation}
where $\Delta_i =-\frac{1}{2} \kappa_i \mu_N B$,
where, $\kappa_i$, is as defined in the electromagnetic tensor term
in the Lagrangian density given by (\ref{lmag}).
For given value of the baryon density, $\rho_B$, and the isospin 
asymmetry parameter, $\eta= ({\rho_n -\rho_p})/({2 \rho_B})$,
the values of the scalar fields
are solved from the equations (\ref{sigma}), (\ref{zeta}), (\ref{delta})
and (\ref{chi}).

The medium modifications of the masses of the bottomonium states  
in the nuclear medium arise due to the modification of the
gluon condensates, which is calculated within the chiral SU(3) model,
from the medium change of the expectation value of the dilaton field.
Equating the trace of the energy momentum tensor of QCD
in the massless quarks limit \cite{cohen}, using the one loop beta
function, with $N_c=3$ and $N_f$=3, to that of the chiral 
effective model as used in the present work,
leads to the relation of the dilaton field to the scalar gluon condensate
as given by \cite{charmonium_mag}
\begin{equation}
\left\langle  \frac{\alpha_{s}}{\pi} G_{\mu\nu}^{a} G^{ \mu\nu a} 
\right\rangle =  \frac{8}{9}(1 - d) \chi^{4}
\label{chiglu}
\end{equation}

The mass shift of the bottomonium state 
arises due to the medium modification of the
scalar gluon condensate, and hence due to the change 
in the value of the dilaton field, and is given as
\cite{amarvdmesonTprc,amarvepja}
\begin{equation}
\Delta m_{\Upsilon}= \frac{4}{81} (1 - d) \int dk^{2} 
\langle \vert \frac{\partial \psi (\vec k)}{\partial {\vec k}} 
\vert^{2} \rangle
\frac{k}{k^{2} / m_{b} + \epsilon}  \left( \chi^{4} - {\chi_0}^{4}\right), 
\label{massupsln}
\end{equation}
where 
\begin{equation}
\langle \vert \frac{\partial \psi (\vec k)}{\partial {\vec k}} 
\vert^{2} \rangle
=\frac {1}{4\pi}\int 
\vert \frac{\partial \psi (\vec k)}{\partial {\vec k}} \vert^{2}
d\Omega,
\end{equation}
$m_b$ is the mass of the bottom quark, taken as 5.36 GeV,
$m_\Upsilon$ is the vacuum mass of the bottomonium state 
and $\epsilon = 2 m_{b} - m_{\Upsilon}$. 
The values of the dilaton field
in the nuclear medium and in the vacuum are $\chi$ and $\chi_0$  
respectively.
$\psi (\vec k)$ is the wave function of the bottomonium state
in the momentum space, normalized as $\int\frac{d^{3}k}{(2\pi)^{3}} 
\vert \psi(\vec k) \vert^{2} = 1 $ \cite{leetemp}.
The wave functions of the bottomonium states can be obtained 
using Fourier transformations of the wave functions
in the co-ordinate space, which are assumed to be
harmonic oscillator wave functions \cite{friman}
and are given as 
\begin{equation}
\psi_{Nlm} (\vec r) = N_{N l}
\times 
(\beta^{2} r^{2})^{\frac{1}2{} l} \exp\Big({-\frac{1}{2} \beta^{2} r^{2}}
\Big) 
L_{N - 1}^{l + \frac{1}{2}} \left( \beta^{2} r^{2}\right)
Y_{lm} (\theta, \phi) 
\equiv R_{Nl} (r) Y_{lm}(\theta,\phi)
\label{wavefn} 
\end{equation} 
where $\beta = \sqrt {M \omega / h}$ characterizes the strength of the 
harmonic potential, $M = m_{b}/2$ is the reduced mass of 
bottom quark - bottom antiquark system, 
$L_p^ q \left( \beta^{2} r^{2}\right)$
is the associated Laguerre Polynomial.
The wave functions in the co-ordinate space are
normalized as 
\begin{equation}
\int d^3 r |\psi _{Nlm} (\vec r) |^2 =1,
\end{equation}
with 
$\int _0 ^\infty |R_{Nl} (r)|^2 r^2 dr=1$
determining the normalization constants $N_{Nl}$,
and, $Y_{lm} (\theta,\phi)$ are the spherical harmonics satisfying the
orthonormality condition 
$\int Y_{lm}(\theta,\phi)Y_{l'm'}(\theta,\phi)d\Omega 
=\delta_{ll'}\delta_{m m'}$.
The strengths of the harmonic oscillator potential, $\beta$,
for the bottomonium states $\Upsilon (NS), N=1,2,3,4$
are calculated from their observed leptonic decay widths,
$\Upsilon (NS)\rightarrow e^+e^-$. The expression for
the decay width, $\Gamma (\Upsilon(NS) \rightarrow e^+e^-$)
is given as 
\cite{AM_DP_upsilon,amspm_upsilon,repko}
\begin{equation}
\Gamma (\Upsilon(NS) \rightarrow e^+e^-)
=
\frac{4 \alpha^2}{9 m_{\Upsilon (NS)}^2} 
|R_{NS}(r=0)|^2,
\end{equation}
where $\alpha=1/137$, $m_{\Upsilon (NS)}$ is the 
vacuum mass of the bottomonium state $\Upsilon(NS)$,
and $R_{NS}(r)$ is the radial part of the 
wave function for this state.
The values of $\beta$ calculated 
are 1309.2, 915.4, 779.75 and 638.6 MeV
fitted from the observed leptonic decay widths 
of 1.34, 0.612, 0.443 and 0.272 keV
for the $\Upsilon(1S)$,
$\Upsilon(2S)$, $\Upsilon(3S)$, $\Upsilon(4S)$,
respectively.  
The vacuum masses of the states, $\Upsilon (1S)$, 
$\Upsilon(2S)$, $\Upsilon(3S)$, $\Upsilon(4S)$,
are 9460.3, 10023.26, 10355.2 and 10579.4 MeV
respectively. The $\beta$ values as 
stated above for these states are observed 
to be smaller for larger masses of the upsilon state.
We use the value of the harmonic oscillator
potential strength parameter, $\beta$
for $\Upsilon (1D)$ (vacuum mass of 10163.7 MeV)
as 858 MeV, which is the value obtained from interpolation 
from the plot of mass versus $\beta$ for the upsilon states.
The leptonic decay width for $\Upsilon (1D)$ can be
calculated by using the formula 
\cite{repko}
\begin{equation}
\Gamma (\Upsilon(ND) \rightarrow e^+e^-)
= \frac{25 \alpha^2}{18 m_{\Upsilon (ND)}^2 m_b^4} 
|R^{''}_{ND}(r=0)|^2.
\end{equation}
With the value of $\beta$ as 858 MeV for $\Upsilon(1D)$,
the decay width 
$\Gamma (\Upsilon(1D) \rightarrow e^+e^-)$
is found to be 0.00715 keV. This value is very small, 
as compared to the values for the leptonic decay widths 
of the $\Upsilon (NS)$ states. The value for
$\Gamma (\Upsilon(1D) \rightarrow e^+e^-)$
obtained in the present study may be compared with
the value of 0.02 keV calculated in Ref. \cite{repko}
using a potential model for study of the heavy quarkonium
states.

In the next section we shall present the results obtained 
for the in-medium upsilon masses in asymmetric nuclear matter 
in presence of strong magnetic fields.

\section{Results and Discussions}

In this section, we first investigate the effects of
magnetic field, density and isospin asymmetry of the magnetized
nuclear medium on the dilaton field $\chi$, which mimics
the gluon condensates of QCD, within the chiral SU(3) 
model. The in-medium masses of upsilon states 
($\Upsilon (NS), N=1,2,3,4,\; \Upsilon (1D)$),
are then calculated from 
the value of $\chi$ in the nuclear medium using 
equation (\ref{massupsln}). 

\begin{figure}
\includegraphics[width=16cm,height=16cm]{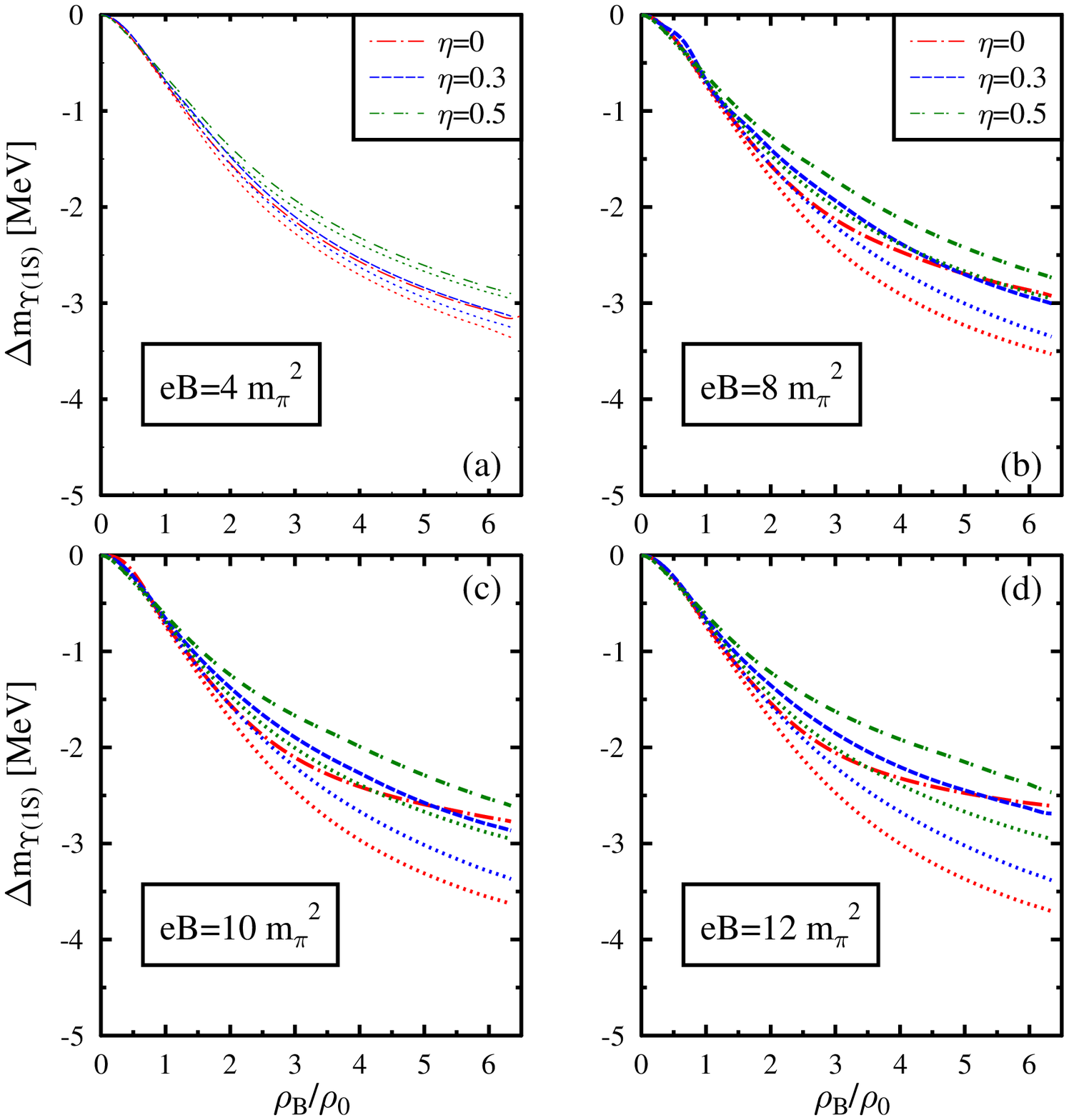}
\caption{(Color online)
The mass shift of $\Upsilon (1S)$ plotted as a function of the
baryon density in units of nuclear matter saturation density,
for different values of the magnetic field and isospin 
asymmetry parameter, $\eta$, including the effects of the
anomalous magnetic moments of the nucleons. The results
are compared to the case when the effects of anomalous magnetic 
moments are not taken into consideration (shown as dotted lines).
}
\label{mupsln_1s_mag}
\end{figure}

\begin{figure}
\includegraphics[width=16cm,height=16cm]{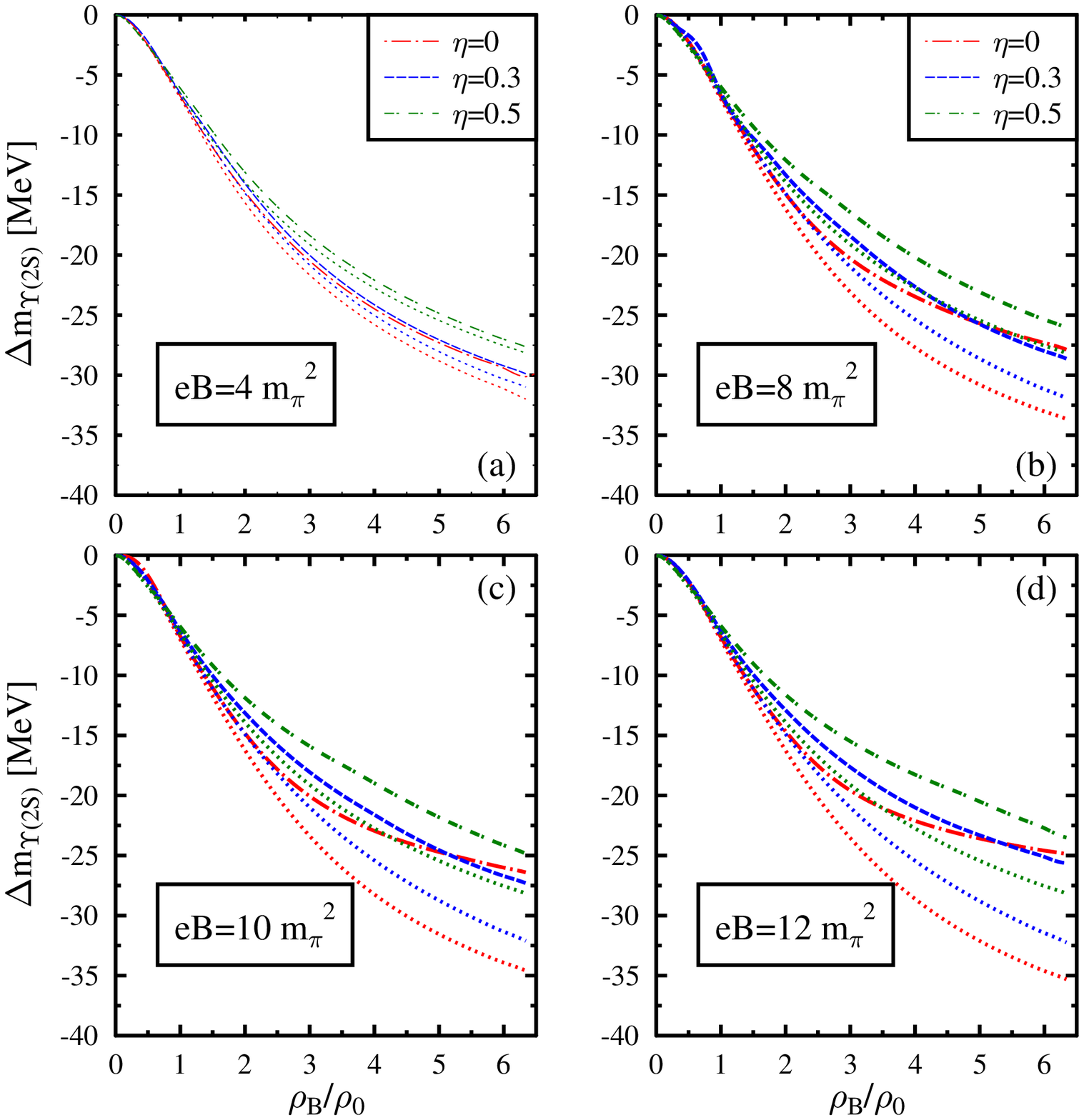}
\caption{(Color online)
The mass shift of $\Upsilon (2S)$ plotted as a function of the
baryon density in units of nuclear matter saturation density,
for different values of the magnetic field and isospin 
asymmetry parameter, $\eta$, including the effects of the
anomalous magnetic moments of the nucleons. The results
are compared to the case when the effects of anomalous magnetic 
moments are not taken into consideration (shown as dotted lines).
}
\label{mupsln_2s_mag}
\end{figure}

\begin{figure}
\includegraphics[width=16cm,height=16cm]{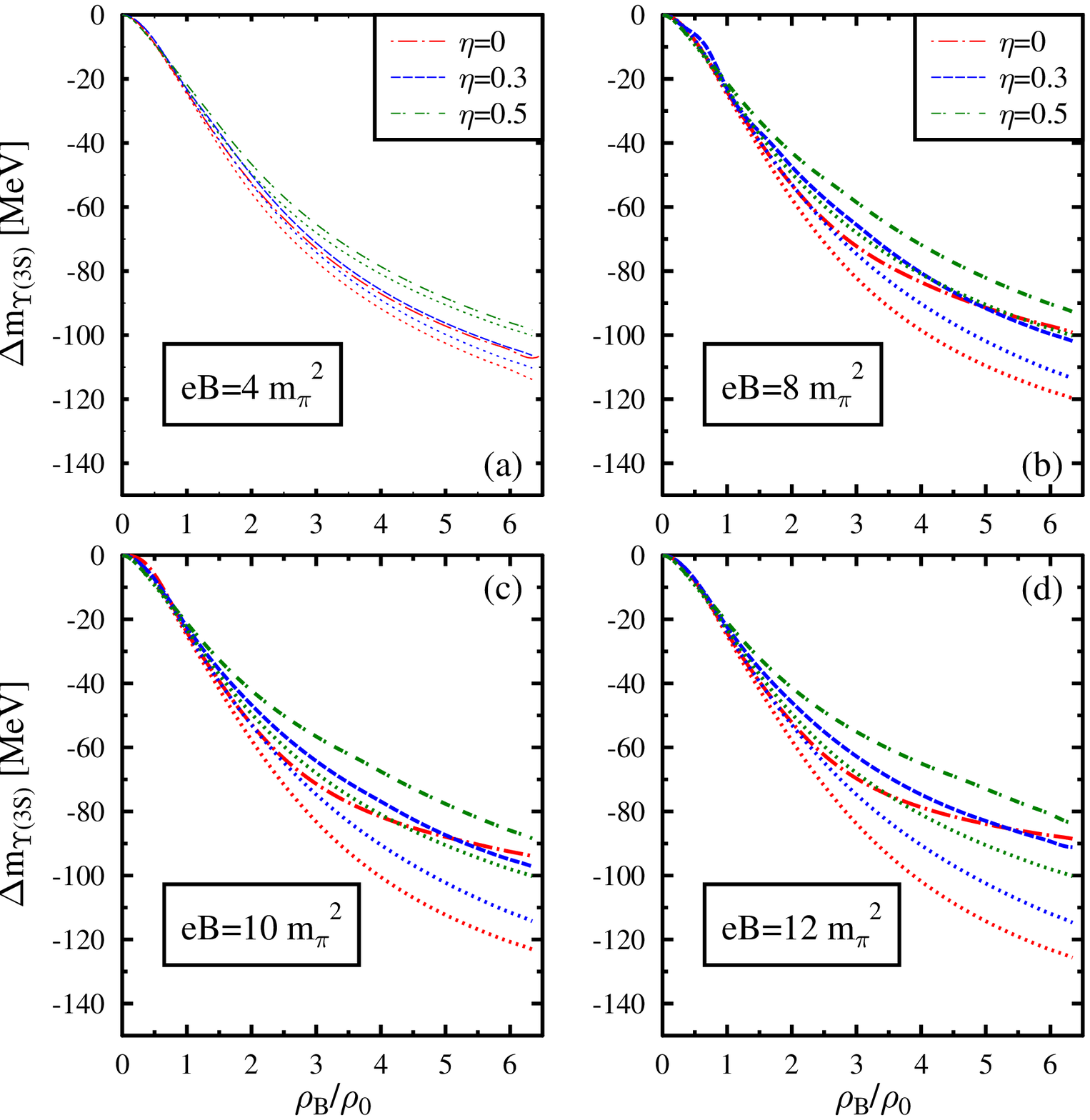}
\caption{(Color online)
The mass shift of $\Upsilon (3S)$ plotted as a function of the
baryon density in units of nuclear matter saturation density,
for different values of the magnetic field and isospin 
asymmetry parameter, $\eta$, including the effects of the
anomalous magnetic moments of the nucleons. The results
are compared to the case when the effects of anomalous magnetic 
moments are not taken into consideration (shown as dotted lines).
}
\label{mupsln_3s_mag}
\end{figure}

\begin{figure}
\includegraphics[width=16cm,height=16cm]{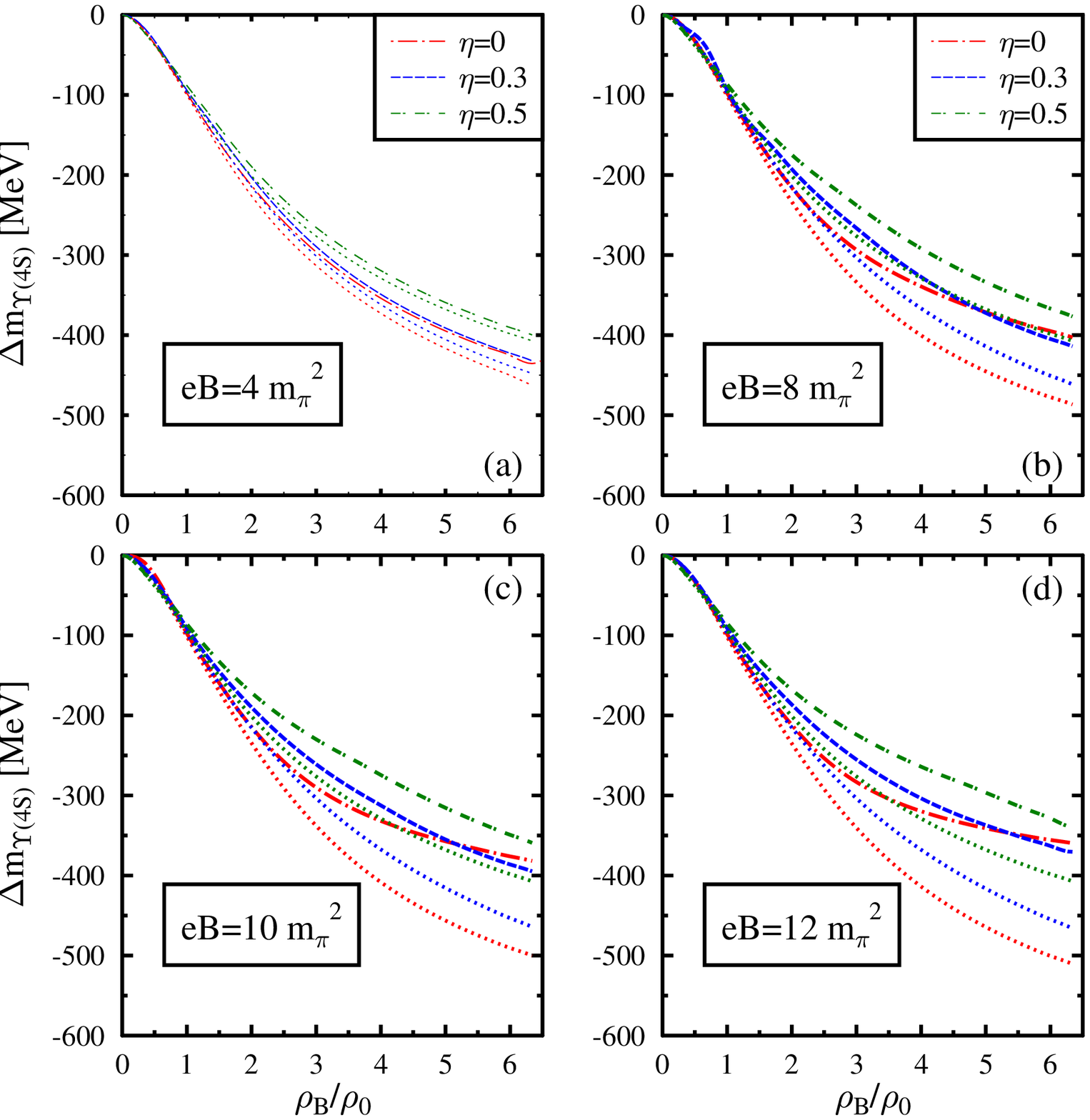}
\caption{(Color online)
The mass shift of $\Upsilon (4S)$ plotted as a function of the
baryon density in units of nuclear matter saturation density,
for different values of the magnetic field and isospin 
asymmetry parameter, $\eta$, including the effects of the
anomalous magnetic moments of the nucleons. The results
are compared to the case when the effects of anomalous magnetic 
moments are not taken into consideration (shown as dotted lines).
}
\label{mupsln_4s_mag}
\end{figure}

\begin{figure}
\includegraphics[width=16cm,height=16cm]{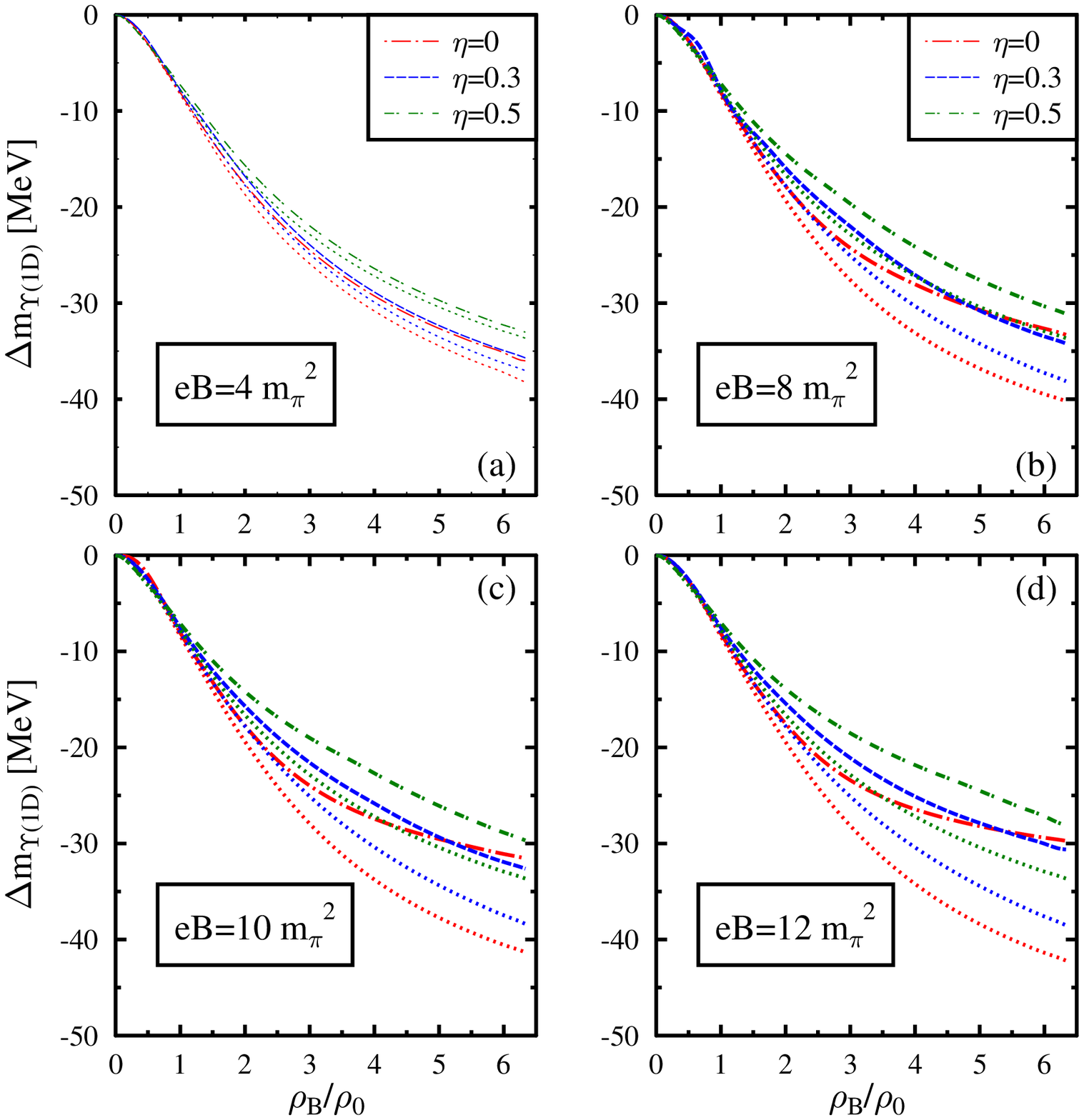}
\caption{(Color online)
The mass shift of $\Upsilon (1D)$ plotted as a function of the
baryon density in units of nuclear matter saturation density,
for different values of the magnetic field and isospin 
asymmetry parameter, $\eta$, including the effects of the
anomalous magnetic moments of the nucleons. The results
are compared to the case when the effects of anomalous magnetic 
moments are not taken into consideration (shown as dotted lines).
}
\label{mupsln_1d_mag}
\end{figure}

The dilaton field $\chi$ as modified
in the asymmetric nuclear medium in the presence of strong magnetic fields,
has been discussed in detail in Ref. \cite{charmonium_mag}.
The variations of the dilaton field $\chi$ with magnetic field,
baryon density, and isospin asymmetry, within the chiral SU(3) model, 
are obtained by solving the coupled equations of motion of the
scalar fields, $\sigma$, $\zeta$, $\delta$ and $\chi$. 
The number density and scalar density of the proton have contributions
from the Landau energy levels in the presence of the magnetic field.
The value of $\chi$ is observed to decrease with increase in density.
When the anomalous magnetic moments (AMM) of the nucleons 
are taken into consideration, at a given density, with increase 
in magnetic field, $\chi$ is observed to attain a higher value 
and thus the shift from the vacuum value decreases. 
The mass shifts from vacuum value, on the other hand,
are observed to increase with increasing magnetic field, 
when the anomalous magnetic moments of the nucleons are not
taken into account. 
As the isospin asymmetry of the medium increases, 
the scalar field $\chi$ is also observed to increase.
For the case of $\eta$=0.5, the medium comprises of
only neutrons, and hence the only effect of magnetic field
is due to the anomalous magnetic moment of the neutrons.
Hence, in the case when the AMM effects are not taken into
consideration, the value of the dilaton field remains
independent of the magnetic field. 
This leads to the mass shift of the upsilon states
to be independent of the magnetic field for $\eta$=0.5, 
when AMM effects are not taken into account.

The masses of the bottomonium states, $\Upsilon (1S)$,
 $\Upsilon (2S)$, $\Upsilon (3S)$, $\Upsilon (4S)$, and $\Upsilon (1D)$,
in magnetized nuclear matter, as calculated from the medium change 
of the dilaton field, are plotted as functions
of the baryon density (in units of nuclear matter
saturation density) in figures \ref{mupsln_1s_mag},
\ref{mupsln_2s_mag},\ref{mupsln_3s_mag},\ref{mupsln_4s_mag},
and, \ref{mupsln_1d_mag} respectively. These
are plotted for values of $eB$ as $4m_\pi^2$, $8m_\pi^2$, 
$10m_\pi^2$, and, $12m_\pi^2$. 
The masses of the bottomonium states are illustrated 
for the values of the isospin asymmetry parameter 
as $\eta$=0, 0.3 and 0.5, with (without) accounting
for the anomalous magnetic moments (AMM) of the nucleons.

The mass modifications of bottomonium states are found to be larger 
in symmetric nuclear matter than in asymmetric nuclear matter. 
As value of $\eta$ increases the mass drop decreases. For a particular 
magnetic field, at a fixed density the effect of anomalous magnetic 
moments results in smaller mass modifications for the bottomonium states 
in comparison to the case when we do not incorporate anomalous magnetic 
moment effects. 

For a fixed value of $\eta$, when we account for the effects of anomalous 
magnetic moments of the nucleons, as magnetic field increases, 
the mass modifications 
of bottomonium states mostly remain constant with negligible variation 
at lower densities. In the same case at higher densities, the mass drop 
decreases as magnetic field becomes larger. 
On the other hand, when AMM effects are not taken into account, 
in symmetric nuclear matter, the mass drop 
increases at higher densities and at larger magnetic fields. 
For asymmetric nuclear matter, without AMM, the mass modifications 
of bottomonium states at a fixed density remains mostly independent 
of magnetic field. Particularly for the case of $\eta$ =0.5 when 
the medium is devoid of protons, magnetic field has no effect 
on mass modifications, since neutrons respond to the magnetic 
field only due to their anomalous magnetic moment.

At a given magnetic field, as baryon density increases, 
there is a general drop in the value of $\chi$ and hence in general 
it is found that the mass shifts of all bottomonium states 
steadily increase as functions of density at a fixed value 
of magnetic field. As density increases the effect of magnetic field 
is more prominent. It has to be noted that, for a given density, 
magnetic field, and, isospin symmetry, the ratio 
of the magnitudes of the mass shifts for the bottomonium states 
turns out to be the ratio of the magnitudes of the integrals
(given in equation (\ref{massupsln}), 
calculated from their respective wave functions in momentum
space. The mass drop of $\gamma(1S)$ turns out to be extremely small 
whereas  $\Upsilon(1D)$, $\Upsilon(2S)$, $\Upsilon(3S)$ are
observed to have larger mass drops, the largest being for
$\Upsilon (4S)$. 

For symmetric nuclear matter ($\eta$ =0), at $\rho_B$ =$\rho_0$(4$\rho_0$), 
including the effects of AMM, the mass shifts (in MeV) of $\Upsilon(1S)$ 
are obtained as $-0.716 (-2.57)$, $-0.7163(-2.463)$, $-0.7145(-2.41)$ and 
$-0.722 (-2.32)$ for magnetic fields $4m_\pi^2$, $8m_\pi^2$,
$10m_\pi^2$, and $12m_\pi^2$ respectively. Under the same conditions, 
the mass shifts of $\Upsilon(1D)$ are found to be $-8.09 (-29.254)$, 
$-8.09 (-28.04)$, $-8.06(-27.43)$, $-8.149(-26.435)$ and those  
of $\gamma(2S)$ is obtained as $-6.82 (-24.494)$, $-6.826 (-23.475)$, 
$-6.81 (-22.96)$ and $-6.88 (-22.12)$. The mass shifts of $\Upsilon(3S)$ 
are $-24.28 (-87.15)$, $-24.29 (-83.53)$, $-24.23 (-81.7)$ 
and $-24.47 (-78.73)$ and of $\Upsilon(4S)$, these have values of 
$-98.66 (-354.17)$, $-98.7 (-339.44)$, $-98.45 (-332)$ 
and $-99.46 (-319.9)$ respectively for the corresponding magnetic fields.
The effects of magnetic feild on the in-medium upsilon masses
are observed to be marginal for small densities, upto around 
nuclear matter saturation density, whereas the effects are observed
to be larger at higher densities. The effects of the anomalous
magnetic moments (AMM) are seen to be appreciable at higher
densities for larger values of the magnetic field.
The mass shifts of $\Upsilon(1S)$,  $\Upsilon(1D)$,  $\Upsilon(2S)$, 
$\Upsilon(3S)$, $\Upsilon(4S)$, at density $4\rho_0$,
with AMM effects in symmetric nuclear matter,
as stated above, may be compared with the values of
$-2.707, -37.22, -25.8, -91.82$, and $-373.13$ 
for $eB=4m_\pi^2$ 
and $-3, -34.21,  -28.64, -101.92$ and $-414.17$ for 
for $eB= 12m_\pi^2$, when the effects from AMM are not taken
into account. 

%
For asymmetric nuclear matter ($\eta$ =0.3), 
at $\rho_B$ =$\rho_0$(4$\rho_0$), including the effects of AMM, 
the mass shifts (in MeV) of $\Upsilon(1S)$, $\Upsilon(1D)$, 
$\Upsilon(2S)$, $\Upsilon(3S)$, $\Upsilon(4S)$,
are observed to be $-0.688 (-2.535)$,  
$-7.77 (-28.85)$, $-6.56 (-24.15)$, $-23.33 (-85.95)$,
and $-94.79 (-349.26)$ for the $eB=4m_ \pi^2$ 
and, $-0.666 ( -2.206)$, $-7.52 (-25.1)$,
$ -6.35 (-21.02)$, $-22.59 (-74.79)$, and,  $-91.78 (-303.92)$ 
for the $eB=12m_\pi^2$. 

\section{Summary}

The medium modifications of the masses of the bottomonium states 
($\Upsilon(1S)$, $\Upsilon(2S)$, $\Upsilon(3S)$, $\Upsilon(4S)$,
$\Upsilon (1D)$)) in strongly magnetized hadronic matter are 
investigated using a chiral effective model. The variation 
in the masses of bottomonium states are due to the modification 
of dilaton field with change in baryon density, isospin asymmetry
as well as magnetic field. The in-medium upsilon masses
are studied accounting for the anomalous magnetic moments
of the nucleons and are compared to the case when these
effects are not taken into consideration.
The modifications due to magnetic field
are observed to be rather small 
at low baryonic densities.
The effects of magnetic field and isospin asymmetry in causing 
the mass modifications are significant at high densities, 
with excited states of bottomonia showing larger mass drop 
than the ground state. Anomalous magnetic moment effects
are observed to be more pronounced with increase
in isospin asymmetry of the medium, and, are larger
at higher densities. The density effects 
are found to be the dominant medium effects, as compared
to the effects due to isospin asymmetry and the magnetic field. 

\acknowledgements
One of the authors (AM) is grateful to ITP, University of Frankfurt,
for warm hospitality and 
acknowledges financial support from Alexander von Humboldt Stiftung 
when this work was initiated. 
Amal Jahan CS acknowledges the support
towards this work from Department of Science and Technology, 
Govt of India, via INSPIRE fellowship 
(Ref. No. DST/INSPIRE/03/2016/003555).


\end{document}